\documentclass{WileyMSP-template}
\usepackage{physics}
\usepackage{bbold} 
\usepackage[labelfont=bf,width=1\linewidth]{caption}
\usepackage{color} 
\usepackage[normalem]{ulem}

\newcommand{\ak}{\hat{a}^{\dagger}}
\newcommand{\av}{\hat{a}^{\phantom{\dagger}}}

\newcommand{\ssp}{\sigma,\sigma'}

\newcommand{\Op}[1]{\hat{\mathrm{#1}}}
\newcommand{\Ham}{\Op{H}}

\newcommand*{\E}{\mathrm{e}}

\DeclareMathOperator\arctanh{arctanh}

\begin{document}

\pagestyle{fancy}
\rhead{\includegraphics[width=2.5cm]{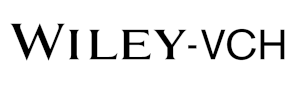}}

\title{Corrections of Electron-Phonon coupling for Second-Order Structural Phase Transitions}

\maketitle


\author{Mario Graml}
\author{Kurt Hingerl}



\begin{affiliations}
DI Mario Graml, Prof. Dr. Kurt Hingerl\\
ZONA, Johannes Kepler Universit\"at, A 4040 Linz, Austria\\
Email Address: mario.graml@jku.at, kurt.hingerl@jku.at

\end{affiliations}



\begin{abstract}

Structural phase transitions are accompanied by a movement of one nucleus (or a few) in the crystallographic unit cell. If the nucleus movement is continuous, a second order phase transition without latent heat results, whereas an abrupt nucleus displacement indicates a first order phase transition with accompanying latent heat. In this paper an Hamiltonian including electron-phonon coupling (EPC) as proposed by Kristoffel and Konsel\cite{Kristoffel1973} is taken. Contrary to their treatment, both the kinetic energy of the nucleus and its position are treated. The interaction of the many electron system with the single nucleus is taken into account by the Born-Oppenheimer approximation and perturbative expressions for the free energies are derived. The nuclei corrections due to the entangled electrons are found to be minor, but highlight the importance of the symmetry breaking at low temperature. Furthermore the free energy for a canonical ensemble is computed, whereas Kristoffel and Konsel used a grand canonical ensemble, which allows to derive more stringent bounds on the free energy. For the zero-order nucleus correction the shift of the phase transition temperature by evaluating the free energy is deduced.

\end{abstract}


\section{Introduction} \label{introduction}

Phase transitions (PT) between two phases $a,b$ are described through their Gibbs energies $G_{a,b}(p,T,H ...)$ at pressure $p$, temperature $T$,  external magnetic field $H$, etc. The Gibbs energy cannot be discontinuous, because all differentials represent intensive quantities. The phase transition temperature (PTT), is then found at the intersection of the hyperplanes in these variables. Despite $G$ has to be continuous also at PTT,  certain derivatives of G can and will be discontinuous. 

For solid-solid or solid-liquid phase transitions with negligible volume change, the free energy $F$ can also be used. Solid-solid PTs occur due to a a change of balance between the binding energy and entropic contributions of different crystal structures. In solid semiconductors, the binding energy is primarily determined by the electronic configuration, expressed through the inner energies $U_{a,b}$, while the entropy contributions $S_{a,b}$ to $F$ arise mainly from low phonon frequencies and phonon dispersion curves as outlined in the chapter by J. Friedel in ref.\cite{Nato}. Even without coupling electronic and nuclear degrees of freedom, first-order phase transitions occur, and the phase transition temperature can be computed given that the binding energy and phonon dispersion for both phases are measured or calculated e.g. using density functional theory. 

However, from a conceptual standpoint, this approach is not satisfactory because the same {\it{ab-initio} Hamiltonian operator describes both phases.} For a deeper understanding, a physical interaction that changes or breaks the symmetry at the phase transition should be included. Hamilton operators that include electron-phonon coupling are well suited for such an approach because when computing $F$, the balance between electronic and phononic contributions varies and both free energies $F_{a,b}$ change as a function of temperature. As a function of temperature the minimum $F(T,p,H)$ is realized, indicating the PT. Landau derived a heuristic theory of phase transitions \cite[446 ff.]{Landau}, while Ginzburg, Anderson, and Cochran connected the occurrence of phase transition with lattice dynamics and electron-phonon interactions \cite{ginzburg1949teoria, I1, I2, I3}. This approach is called "vibronic theory"\cite{vibronicth} and relies on non-analyticity or discontinuity in the derivatives of $F$. At different temperatures, different - continuous or discontinuous- displacements of the nucleus result, indicating first or higher order phase transitions. Kristoffel and Konsel (KK) took up Ginzburg's idea and treated it analytically \cite{Kristoffel1973}, but their approach did not take into account the kinetic energy of the nucleus. Herein we reanalyze and assess the KK results, correct typographical errors, and provide a comprehensive thermodynamic treatment taking into account the nucleus kinetic energy. Our findings demonstrate the crucial role of these corrections in facilitating the formation of entangled electron-phonon states.

Section \ref{Main results of the Kristoffel and Konsel approach} provides a comprehensive review and summary of the Kristoffel and Konsel (KK) approach. The main result of section \ref{Electronic energies} are the electronic energies and wavefunctions; the formal derivation can be found in an appendix. The partition function and the free energy for the KK model, with different number of electrons per unit cell, but not yet including the nucleus correction are computed in section \ref{Helmholtz without Nuclei}. In section \ref{Nuclei_Corrections}, the effect of the kinetic energy of the nucleus for $F$ is addressed. Finally, the key findings and discrepancies compared to the Kristoffel and Konsel approach are summarized in section \ref{Conclusion}.

 \section{Concept of Vibronic Theory}  \label{Main results of the Kristoffel and Konsel approach}

In semiconductors with two bands, an additional term describing the electron-phonon interaction (EPI) is included in the Hamiltonian, and the total (electronic and phononic and electron-phonon interaction) inner energy is calculated employing different approximations using a parametrized vibronic normal coordinate labeled "$X$" (in the QM operator version "$\Op{X}$").  The model Hamiltonian including EPI consists of different contributions: \cite{vibronicth}
\begin{align}
\hspace{-10pt}\Ham &=\underbrace{\sum\limits_{\vec{q}j}\left (\frac{1}{2M_j}\Op{P}_{\vec{q}j}\Op{P}_{-\vec{q}j}+\frac{M_j}{2}\omega_{\vec{q}j}\Op{X}_{\vec{q}j}\Op{X}_{-\vec{q}j}\right )+\sum \limits_{\vec{q}_1j_1} \sum \limits_{\vec{q}_2j_2} \sum \limits_{\vec{q}_3j_3} \sum \limits_{\vec{q}_4j_4} B(\vec{q}_1j_1,\vec{q}_2j_2,\vec{q}_3j_3,\vec{q}_4j_4)\Op{X}_{\vec{q}_1j_1}\Op{X}_{\vec{q}_2j_2}\Op{X}_{\vec{q}_3j_3}\Op{X}_{\vec{q}_4j_4}+....}_{\Ham_{\text{ph}}}\notag \\ 
\hspace{-40pt}&+\underbrace{\sum \limits_{\sigma, \vec{k}}\varepsilon_{\sigma}( \vec{k}) \ak_{\sigma \vec{k}}\av_{\sigma \vec{k}}+\Ham_{\text{el-el}}}_{\Ham_{\text{el}}} +\underbrace{\frac{1}{\sqrt{N}} \sum\limits_{\ssp}\sum\limits_{\vec{k},\vec{k}'}\sum\limits_{\vec{q}j}V_{\sigma \sigma'}^{j} \left (\vec{q}, \vec{k}, \vec{k}' \right )\ak_{\sigma \vec{k}}\av_{\sigma' \vec{k}'}\delta \left (\vec{k}'-\vec{k}+\vec{q} \right )\Op{X}_{\vec{q}j}}_{\Ham_{\text{ph-el}}} \label{ep-el:hamstart}
\end{align} 
The first two summations labeled as $\Ham_{\text{ph}}$ represent the kinetic energy, harmonic and anharmonic potential of the nuclei, while the first two terms in the second line denote the pure electronic contribution (labeled with $\Ham_{\text{el}}$). The last term, labeled as $\Ham_{\text{ph-el}}$, describes the electron-phonon coupling in linear order, which models the vibronic interaction. 

This Hamiltonian Equation \eqref{ep-el:hamstart} serves as our starting point for approximations. In $\Ham_{\text{el}}$ the electronic bands $\varepsilon_{\sigma}(\vec{k})$ are represented depending on the band $\sigma$ and the wave vector $\vec{k}$.  The electron-electron interaction is represented by $\Ham_{\text{el-el}}$. The second quantization is used to write $\Ham_{\text{el}}$ with the electronic creation/destruction operators $\ak/\av$. In $\Ham_{\text{ph}}$ normal coordinate operators $\Op{X}_{\vec{q}j}$ and conjugated momenta $\Op{P}_{\vec{q}j}$ of the vibrations with index $\vec{q}$ appear, representing the wave vector in the vibrational branch $j$. $M_j$ corresponds to the reduced mass of the active vibration with the bare frequency $\omega_{\vec{q}j}$, while $B$ is the coefficient of the fourth-order phonon anharmonicity.  The coupling of electrons and phonons is parametrised by the linear coupling parameter $V_{\sigma \sigma'}^{j} \left ( \vec{q}, \vec{k}, \vec{k}' \right )$ which can be calculated using density functional perturbation theory i.e. Giustino \cite{Giustino2017}. \newline  

We adopt the two-band model developed by Kristoffel et al. \cite{Kristoffel1973} as the basis for our analysis. This model involves several key approximations:
\begin{itemize}
    \item We restrict the Hamiltonian to two dispersionless electronic bands, labeled as $\sigma=1$ and $\sigma=2$, neglecting $\Op{H}_{\text{el-el}}$. These bands have energies denoted by $\varepsilon_{\sigma}$, and exhibit opposite parity. The dispersionless energies are chosen, because a) we are interested mainly in transition close to the band gap, and b) because phonon energies are much smaller than electronic energies.
    \item The band gap between the two bands is  characterized by $\Delta=\varepsilon_2-\varepsilon_1$.
    \item Anharmonicity effects of vibrations are neglected, setting $B=0$.
    \item It is assumed, that only inter-band electron-phonon interaction contributes, i.e. both intra-band electron phonon scattering probabilities are equal and set to zero: $V_{11}=V_{22}=0$. The inter-band interaction $V_{12}$ and $V_{21}$ are denoted by $V$.
\end{itemize}
Based on these approximations, we obtain the two-band Hamiltonian 
\begin{align}
\Ham=&\sum_{\sigma}\varepsilon_{\sigma}\left(\underbrace{\ak_{\sigma}\av_{\sigma}\otimes\mathbb{1}_{H_1}\otimes\mathbb{1}_{H_1}\ldots\otimes\mathbb{1}_{H_1}}_{N_{\rm e}}\otimes\mathbb{1}_{H_\text{nu}}+\mathbb{1}_{H_1}\otimes\ak_{\sigma}\av_{\sigma}\otimes\mathbb{1}_{H_1}\ldots\otimes\mathbb{1}_{H_1}+\ldots + \mathbb{1}_{H_1}\otimes \ldots \otimes \ak_{\sigma}\av_{\sigma}\right)\notag \\ +&\underbrace{\mathbb{1}_{H_1}\otimes\mathbb{1}_{H_1}\otimes\mathbb{1}_{H_1}\ldots\otimes\mathbb{1}_{H_1}}_{N_{\rm e}}\otimes\left(\frac{\Op{P}^2}{2M}+\frac{M\omega^2}{2}\Op{X}^2\right) \notag \\ +& \sum_{\sigma\neq\sigma'}\frac{V}{\sqrt{N}}\left(\underbrace{\ak_{\sigma}\av_{\sigma'}\otimes\mathbb{1}_{H_1}\otimes\mathbb{1}_{H_1}\ldots\otimes\mathbb{1}_{H_1}}_{N_{\rm e}}+\mathbb{1}_{H_1}\otimes\ak_{\sigma}\av_{\sigma'}\otimes\mathbb{1}_{H_1}\ldots\otimes\mathbb{1}_{H_1}+\ldots + \mathbb{1}_{H_1}\otimes \ldots \otimes \ak_{\sigma}\av_{\sigma'}\right)\otimes \Op{X}
 \label{H_full_tens}
 \end{align}
In order to facilitate the analysis of the Hamiltonian in Equation \eqref{H_full_tens}, a shorthand notation is introduced
\begin{align}
\Ham=\underbrace {\sum^{\hspace{25pt}N_\text{e}}\limits_\sigma \varepsilon_\sigma \ak_\sigma\av_\sigma}_{{\Ham_{\text{el}}}}+\underbrace{\frac{1}{2}\left (\frac{\Op{P}^2}{M}+M\omega^2\Op{X}^2\right )}_{\Ham_{\text{ph}}}+\underbrace{\sum^{\hspace{25pt}N_\text{e}} \limits_{\sigma \neq \sigma'}\frac{V}{\sqrt{N}}\ \ak_{\sigma}\av_{\sigma'}\Op{X}}_{\Ham_{\text{ph-el}}} \label{Ham}
\end{align}
which is also employed by Vainstein\cite{Vainstein} in a similar fashion.
The simplified Hamiltonian in Equation \eqref{Ham} consists of a single phonon term, denoted as $\Ham_{\rm ph}$,  $N_{\rm e}$ non-interacting electrons distributed among the two bands $\sigma_1$ and $\sigma_2$, represented by $\Ham_{\rm el}$, and the term $\Ham_{\rm ph-el}$ showing the interaction between $N_{\rm e}$ electrons and the vibration of the single nucleus. Also this simplified Hamiltonian cannot be solved analytically due to the direct entanglement between the electrons and the nucleus. To proceed, we apply the Born-Oppenheimer approximation and incorporate the correction of the nucleus using the second
Bogoliubov inequality\cite{Tyablikov1967,Bogoliubov-Encyclopedia}.

\section{Electronic Energies arising from Born-Oppenheimer Approximation} \label{Electronic energies}
The objective is to find a solution to the Schrödinger equation for the given Hamiltonian $\Ham\psi=E\psi$, where $E$ is the energy of the coupled system and $\psi$ is the entangled wavefunction, depending on the positions of all the electrons and the nucleus. Due to the significant difference in mass between the ions and the electrons, we make the Born-Oppenheimer (BO) approximation with the following ansatz:

\begin{equation}
\psi(X)=\Phi(X)\chi(X) \label{psi}.
\end{equation}

Equation \eqref{psi} is the product ansatz for the total wavefunction $\psi(X)$ with the electronic wavefunction $\Phi(X)$ and the nuclear wavefunction $\chi(X)$. (We are working in second quantization, in first quantization this is written as $\psi(x_{i},X)=\Phi(x_{i},X)\chi(X)$) Note that the electronic wavefunction $\Phi$ depends on the nuclear coordinate $"X"$ only as a parameter. 

\begin{subequations} \label{BO}
\begin{equation}
\Ham \approx \Ham_{\text{BO}}=\Ham_{\text{el,BO}}+\frac{\Op{P}^2}{2M}
\end{equation}
\begin{equation}
\Ham_{\text{el,BO}}=\sum^{\hspace{25pt}N_\text{e}} \limits_{\sigma} \varepsilon_{\sigma}\ak_{\sigma}\av_{\sigma}+\frac{1}{2}M\omega^2X+\sum^{\hspace{25pt}N_\text{e}} \limits_{\sigma \neq \sigma'}\frac{V}{\sqrt{N}}\ \ak_{\sigma}\av_{\sigma'}X\label{H_el}
\end{equation}
\end{subequations}

The first line of Equation \eqref{BO} separates the motion of the nucleus from the electronic Hamiltonian, while the second line represents the electronic Hamiltonian. To calculate the correction term for the kinetic energy of the nucleus, we derive the electronic energy $\bar{\varepsilon}_{N_1}(X)$ \ref{renormen2} and the associated wavefunction $\Phi_{N_1}(X)$ \ref{renormen1} in Appendix \ref{Appendix_Electronic}. The electronic energy was calculated as follows:
\begin{align}
\bar\varepsilon_{N_1}(X) = N_1\varepsilon'_{+}(X) + (N_\text e - N_1)\varepsilon'_{-}(X) + \frac{M\omega^2}{2}X^2
\end{align}

where $\varepsilon'_{\mp}(X)$ is defined as:

\begin{align}
\varepsilon'_{\pm}(X) = \frac{\varepsilon_1 + \varepsilon_2 \pm \sqrt{\Delta^2 + 4\frac{V^2}{N}X^2}}{2}
\end{align}

The wavefunction was calculated, noting that each electron is entangled with the nucleus

\begin{align}
\Phi_{N_1}(X) = \Phi_{\pm_1}(X) \otimes \Phi_{\pm_2}(X) \otimes \ldots \otimes \Phi_{\pm_{N_e}}(X)
\end{align}

where $\Phi_{\pm}(X)$ is given by:

\begin{align}
\Phi_{\pm}(X) = \pm\frac{1}{\sqrt{2}}\sqrt{1 \pm \frac{\Delta}{E_{\rm g}(X)}}\varphi_{1} + \frac{1}{\sqrt{2}}\sqrt{1 \mp \frac{\Delta}{E_{\rm g}(X)}}\varphi_{2}
\end{align}

where the band-gap with electron-phonon interaction is defined as $E_{\text g}(X) = \varepsilon'_{+}(X) - \varepsilon'_{-}(X)$.
 Based on the above results, the following scalar products can now be calculated. These products are necessary for the kinetic energy correction due to the nucleus.

\begin{subequations}
\begin{align}
\braket{\Phi_{N_1}(X)}{\pdv{\Phi_{N_1}(X)}{X}} = 0 \label{mn1}
\end{align}
\begin{align}
\braket{\Phi_{N_1}(X)}{\pdv[2]{\Phi_{N_1}(X)}{X}} = -\frac{N_\text e}{N}\frac{V^2 \Delta^2}{E_\text g^4(X)} \label{mn2}
\end{align}
\end{subequations}

These scalar products provide valuable insights into the behavior of the electronic system under. The symmetry of the system implies that the first-order derivative in Equation \eqref{mn1} contributes zero, fulfilling time-reversal symmetry\cite[sec. 8]{Grosso2014}. However, the electron-phonon coupling equally affects the significance of the second-order derivative for both bands. After determining the electronic energies, the next step is to calculate the Helmholtz free energy.

\section{Helmholtz Energy without Nuclei-Correction}\label{Helmholtz without Nuclei} 
This chapter presents a derivation of the KK results from a different perspective. As previously stated, the simplest approximation neglects the kinetic energy contribution of the nucleus, resulting in $X$ being treated as a parameter. Therefore, the total energy of the system is approximated as the electronic energy, which is expressed as follows:
\begin{equation}
E_{N_1}(X)\approx \bar{\varepsilon}_{N_1}(X) \label{Eklasisch}
\end{equation}
To determine the phase transition temperature, we must calculate the Helmholtz free energy. The Helmholtz energy, denoted by $F$, is typically obtained using the partition function $Z$:
\begin{align}
F=-\frac{1}{\beta}\ln Z=-\frac{1}{\beta}\ln\left(\Tr\left(\E^{-\beta \Op{H}}\right)\right)\label{F-q-g}
\end{align}
To evaluate the trace in this equation, we use the following expression:
\begin{equation}
\Tr\left(\E^{-\beta\Ham}\right)=\int\limits_{\lambda \in \textit{Spec}_{\Ham}}\E^{-\beta\lambda}\ \dd p_{\Ham}(\lambda)
\end{equation}
where $\dd p_{\Ham}(\lambda)$ represents the measure and $\beta=1/k_\text B T$ the coldness. We previously encountered a challenge in dealing with the continuous spectra of our measure, which requires a more sophisticated approach compared to handling points. To overcome this challenge, we adopt an additional approximation. Since the free energy takes the minimum value, we argue that the minimization of the Helmholtz energy trough varying $X$ is a necessary and sufficient condition. Therefore the position $X$ will operate as relevant order-parameter. This approach is commonly referred to as the "static approximation" \cite{Phd60}. Kristoffel et al. \cite{Kristoffel1973} took a similar approach, using the grand canonical approximation for the canonical ensemble instead of directly calculating the ensemble. However, their results differed significantly from ours. We calculated the Helmholtz energy using the static approximation, which yielded:
\begin{align}
F(N_{\rm e},\beta,X)&=-\frac{1}{\beta}\ln \left(\sum\limits_{N_1=0}^{N_\text e}\binom{N_\text e}{N_1}\E^{-\beta N_1\varepsilon'_+(X)}\E^{-\beta (N_\text e-N_1)\varepsilon'_-(X)}\E^{-\beta\frac{M\omega^2}{2}X^2} \right ) \notag \\ &=N_{\rm e} \frac{\varepsilon_1+\varepsilon_2}{2}-\frac{N_{\rm e}}{\beta} \ln\left (2\cosh\left(\frac{\beta E_{\text g}(X)}{2} \right ) \right )+ \frac{M \omega^2}{2}X^2 \label{full-energy}
\end{align}
The Helmholtz energy $F$ is composed of three terms. The first term represents the mean value of the electronic bands, the third term represents the energy of the lattice, and the second term, which depends on temperature, represents the electron-lattice interaction due to electron-phonon coupling. In the static approximation, the minimization of the Helmholtz energy with respect to $X$ requires the following two conditions for real $X_0$:
\begin{subequations}
\begin{equation}
\eval{\pdv{F}{X}}_{X_0}=0 \label{condi1}
\end{equation}
\begin{equation}
\eval{\pdv[2]{F}{X}}_{X_0}>0 \label{condi2}
\end{equation}
\end{subequations}
Evaluating and reformulating Equation \eqref{condi1} one finds 
\begin{equation}
\frac{M\omega^2}{2V^2}\frac{N}{N_{\rm e}}X_0=\frac{\tanh\left(\frac{\beta E_{\text g}(X_0)}{2}\right) }{E_{\text g}(X_0)}X_0.
\label{confx0}
\end{equation}
Equation \eqref{confx0} shows that the solution $X_0=0$ corresponds to the realization of the first phase, whereas an other solution $X_0\neq 0$ indicates the second phase. For $X_0\neq 0$, an implicit Equation \eqref{x0} for $X_0^2(\beta)$ is derived: 
\begin{equation}
 X_0^2(\beta)=\frac{1}{N}\left (\frac{VN_{\rm e}}{M \omega^2}\tanh\left (\frac{\beta E_{\text g}((X_0))}{2} \right )\right )^2-N\left (\frac{\Delta}{2V}\right )^2     \label{x0}
\end{equation}
with the following realizable low temperature and non realizable high-temperature limits:
\begin{align}
& X_0^{2}(\beta \rightarrow \infty)=\frac{1}{N}\left (\frac{VN_{\rm e}}{M \omega^2}\right )^2-N\left (\frac{\Delta}{2V}\right )^2 &   X_0^{2}(0) =-N\left (\frac{\Delta}{2V}\right )^2  \label{x0-limit}  
\end{align}
At low temperatures, we have a maximal lattice distortion, which decreases as the temperature increases. We can obtain $X_0\neq 0$ if 
\begin{align}
&\frac{1}{N}\left (\frac{VN_{\rm e}}{M \omega^2}\right )^2>N\left (\frac{\Delta}{2V}\right )^2& &\longleftrightarrow&  &\frac{M\omega^{2}\Delta}{2V^2}\frac{N}{N_{\rm e}} <1&  \label{condeval1}
\end{align}
is satisfied, which is also the condition found by KK \cite{Kristoffel1973}. What remains to be shown is that condition \eqref{condi2} is a minimum. Evaluating Equation \eqref{condi2}, yields:

\begin{equation}
\pdv[2]{F}{X} = M\omega^2 - \frac{N_{\rm e}}{2}\left(\pdv[2]{E_{\rm g}(X)}{X} \tanh\left(\frac{\beta E_{\rm g }(X)}{2}\right) + \left(\pdv{E_{\rm g}(X)}{X}\right)^2 \frac{\beta}{2}\frac{1}{\cosh^2\left(\frac{\beta E_{\text g}(X)}{2}\right) }\right) \label{evcond2}
\end{equation}

For the solution where  $X_0 \neq 0$, it is not possible to directly derive the minimum. However, the nature of this extremum can be determined by demonstrating that when $X_0=0$ for condition \eqref{condeval1}, $F$ reaches a maximum. As there are only two solutions for $X_0$, the alternative solution $X_0 \neq 0$ must represent a minimum of $F$. Furthermore one finds that a second-order phase transition results with  where two phases coexisting at the PTT, because the second derivative of  $F$ changes sign for at condition \eqref{condeval1}.  The Curie temperature, either denoted as $T_{\text{C}}$ or inverse Curie temperature $\beta_{\text{C}}$, can be calculated by substituting $X_0=0$ into Equation \eqref{x0}:

\begin{equation}
k_{\text B}T_{\text C} = \frac{\Delta}{2}\frac{1}{\arctanh(\tau) } \label{Tc}
\end{equation}

where

\begin{equation}
\tau = \frac{M\omega^{2}\Delta}{2V^2}\frac{N}{N_{\text e}}
\end{equation}

The value of $\tau$ must be between 0 and 1 for it to be a valid argument of arctanh. Equation \eqref{Tc} demonstrates that a high phase transition temperature necessitates strong electron-phonon coupling, a small band gap, and/or a low frequency phonon mode.

\section{Helmholtz Energy with Nucleus-Corrections}\label{Nuclei_Corrections}
To incorporate the corrections of the kinetic energy of the nucleus into the calculation of the free energy, we use the second Bogoliubov inequality\cite{Tyablikov1967,Bogoliubov-Encyclopedia}. As demonstrated by Gidopoulos \cite{Gidopoulos_2022}, the kinetic energy of the nucleus is incorporated via quantum mechanical perturbation theory. The resulting term is referred to as the diagonal Born-Oppenheimer correction. Here, a similar strategy is employed, with the exception that Bogoliubov is utilized instead. In this particular case the correction $F'$ is calculated using the static approximation and Equation \eqref{mn1} as
\begin{equation}
    F'=-\frac{\hbar^2}{2M}\frac{\sum\limits_{\sigma_1\ldots\sigma_{N_{\text e}}}\exp(-\beta \sum\limits_{i=1}^{N_{\text{e}}}\bar{\varepsilon}_{\sigma_i}(X))\left( \beta^2\left( \pdv{\sum\limits_{j=1}^{N_{\text{e}}}\bar{\varepsilon}_{\sigma_{j}}(X)}{X} \right)^2-\beta\pdv[2]{\sum\limits_{m=1}^{N_{\text{e}}}\bar{\varepsilon}_{\sigma_{m}}(X)}{X}+\sum\limits_{n=1}^{N_{\text e}} \braket{\phi_{\sigma_{n}X}}{\pdv[2]{\phi_{\sigma_{n}X}}{X}} \right)}{\left(\sum\limits_{\sigma}\exp(-\beta \bar{\varepsilon}_{\sigma}(X))\right)^{N_{\text e}}}   \label{pertura}
\end{equation}
From Equation \eqref{pertura}, it is evident that the nucleus correction are significant for low temperatures. Upon evaluating Equation \eqref{pertura} and arranging the outcome in terms of increasing powers of $\beta$, one arrives at
\begin{align}
F'(N_\text e,\beta,X)=N_{\text e}\frac{\hbar^2}{2M}\Bigg (\frac{V^2}{N}\left( \frac{\Delta}{E_{\text g}^2(X)} \right)^2&- \frac{\beta}{2}\pdv[2]{E_{\text g}(X)}{X}\tanh\left( \frac{\beta E_{\text g}(X)}{2}  \right)\notag \\&-\frac{\beta^2}{4}\left( \pdv{E_{\text g}(X)}{X} \right)^2 \left( 1+\tanh\left( \frac{\beta E_{\text g}(X)}{2} \right)^2\left( N_{\text e}-1 \right) \right)  \Bigg)\label{narrow band:F'}
  \end{align}
In Equation \eqref{narrow band:F'}, three novel aspects of the corrections are elucidated when contrasted with the unperturbed scenario: Firstly, the corrections exhibit a dependence proportional to the temperature, contrary to the electron system's behavior. Secondly, these corrections now rely on the ratio of $N_{\text{e}}/N$. Lastly, they encompass the influence of band bending. Notably, all nucleus corrections are on the order of $M^{-1}$, implying that for large $M$ and typical temperatures, they remain insignificantly small, as initially assumed. \\
 Including the correction Equation \eqref{narrow band:F'} into Equation \eqref{condi1} one find the nucleus corrected condition:
 
\begin{align}
 \hspace{0pt}\pdv{F}{X}&=X M \omega ^2-\frac{ N_{\text e}}{2}  \tanh \left(\frac{ \beta   E_{\text g}}{2}\right)\pdv{ E_{\text g}}{X}-2\frac{\hbar ^2}{M }\frac{N_{\text e}}{N}\frac{\Delta ^2 V^2}{   E_{\text g}^5}\pdv{ E_{\text g}}{X} \notag \\
 \hspace{0pt}&-\frac{ N_{\text e}\hbar^2}{8M}\Bigg(2\beta \pdv[3]{ E_{\text g}}{X} \tanh \left(\frac{\beta E_{\text g}}{2}\right)+\beta^2\pdv{ E_{\text g}}{X} \pdv[2]{ E_{\text g}}{X} \left(3-(3-2 N_{\text e}) \tanh\left(\frac{ \beta   E_{\text g}}{2}\right)^2\right)
 \notag \\
 &\hspace{55pt}+\beta^3\frac{\left( N_{\text e}-1 \right) \pdv{ E_{\text g}}{X} \tanh \left(\frac{ \beta   E_{\text g}}{2}\right)}{\cosh \left(\frac{ \beta   E_{\text g}}{2}\right)^2}\left( \pdv{ E_{\text g}}{X} \right)^2\Bigg)\label{narrow band:F1}
\end{align}
The calculation of the missing derivatives of the renormalised band gap is as follows:
\begin{subequations}
\begin{equation}
\pdv{  E_\text{g}  }{X}=\frac{4V^2}{N  E_\text{g}  }X \label{narrow band: eg1}
\end{equation}
\begin{equation}
\pdv[2]{  E_\text{g}  }{X}=\frac{4V^2\Delta^2}{N  E_\text{g}  ^3} \label{narrow band: eg2}
\end{equation}
\begin{equation}
\pdv[3]{  E_\text{g}  }{X}=-\frac{48V^4\Delta^2 }{N^2  E_\text{g}  ^5}X\label{narrow band: eg3}
\end{equation}
\end{subequations}
Upon substituting Equation \eqref{narrow band: eg1}, \eqref{narrow band: eg2}, and \eqref{narrow band: eg3} into Equation \eqref{narrow band:F1}, it becomes evident that we can factorize $X$, implying that Equation \eqref{condi1} consistently yields the solution $X_0=0$. To further explore this extremum, we employ condition Equation \eqref{condi2} for $X_0=0$
\begin{align}
\eval{\pdv[2]{F}{X}}_{X=0}=&M \omega^2-2\frac{ N_{\text e}}{ N}\frac{V^2 }{\Delta }\tanh \left(\frac{\beta  \Delta }{2}\right) -8\frac{N_{\text e}}{N^2}\frac{  V^4 \hbar ^2}{\Delta^4 M }+12\beta \frac{\hbar^2}{M}\frac{N_{\text e}}{N^2}\frac{V^4}{\Delta^3}\tanh\left( \frac{\beta \Delta}{2} \right) \notag \\
&-2\beta^2\frac{\hbar ^2}{M}\frac{ N_{\text e}}{N^2}\frac{ V^4}{\Delta^2}\left(3-3 \tanh ^2\left(\frac{\beta  \Delta }{2}\right)+2 N_{\text e} \tanh ^2\left(\frac{\beta  \Delta }{2}\right)\right)
\label{narrow band: krümmung}
\end{align}

Equation \eqref{narrow band: krümmung} provides valuable information that can be extracted by considering the high and low temperature limits. We will begin by examining the high temperature limit $\beta \rightarrow 0$:

\begin{equation}
\eval{\pdv[2]{F}{X}}_{X=0} \sim M \omega^2 -8\frac{N_{\text e}}{N^2}\frac{  V^4 \hbar ^2}{\Delta^4 M } \ \text{as} \ \beta \rightarrow 0 \label{narrow band: highT}
\end{equation}
As shown in Equation \eqref{narrow band: highT}, after incorporating the corrections, the curvature's sign now depends on the given parameters. If

\begin{equation}
M\omega^2> 8\frac{N_{\text e}}{N^2}\frac{  V^4 \hbar ^2}{\Delta^4 M } \label{narrow band: nopolaron}
\end{equation}
 the minimum position is at $X_0=0$, indicating that the nucleus is in the middle of the cell. This is similar to the scenario when the nuclei are not considered. 

The second case
\begin{equation}
M\omega^2< 8\frac{N_{\text e}}{N^2}\frac{  V^4 \hbar ^2}{\Delta^4 M }
\end{equation}
implies a negative curvature, indicating a maximum at $X_0=0$. As $X$ increases, the free energy also increases until a minimum with $X_0\neq 0$ is reached, even at  $T \rightarrow \infty$.

The entangled electron and phonon state that will occur can be interpreted as the formation of a polaron\cite[5 ff.]{haken1976quantum}. As demonstrated, the formation of the polaron is only possible by including the kinetic energy of the nuclei. To incorporate the breakdown of this polaron in the presented theory, one must include phonon-phonon interaction.\\ Next, we consider the low-temperature limit $\beta \rightarrow \infty$ and obtain
\begin{equation}
\eval{\pdv[2]{F}{X}}_{X=0} \sim -4\beta^2\frac{\hbar ^2}{M} \left(\frac{ N_{\text e}}{N} \right)^2\frac{ V^4}{\Delta^2} \ \text{as} \ \beta \rightarrow \infty \label{narrow band: lowT1}
\end{equation}
As shown in Equation \eqref{narrow band: lowT1} the ground state/low temperature solution always has a minimum with $X_0\neq 0$, indicating that the symmetry is always broken when including the kinetic energy of the nuclei. This differs greatly from the condition found in literature \cite{Kristoffel1968,Kristoffel1973} Equation \eqref{condeval1}.\\

Assuming moderate temperatures, we aim to study the initial changes in the electronic system. To achieve this, we will only consider the constant part of the correction $\frac{\hbar^2}{2M}\frac{N_\text e}{N}\frac{V^2 \Delta^2}{E_\text g^4(X)}$ in the realm of the static approximation:
\begin{align}
F(N_{\text e},\beta,X) = N_{\text e} \frac{\varepsilon_1+\varepsilon_2}{2}-\frac{N_{\text e}}{\beta} \ln\left (2\cosh\left(\frac{\beta E_{\text g}(X)}{2} \right ) \right )+ \frac{M \omega^2}{2}X^2+\frac{\hbar^2}{2M}\frac{N_\text e}{N}\frac{V^2 \Delta^2}{E_\text g^4(X)} \label{FxB}
\end{align}
The condition for the phase transition has changed due to the additional term $\frac{\hbar^2}{2M}\frac{N_\text e}{N}\frac{V^2 \Delta^2}{E_\text g^4(X)}$:
\begin{align}
0 < \left (\tau-2\left(\frac{\hbar \omega}{\Delta} \right)^2\frac{1}{N_\text e\tau}\right ) \leq 1 \label{condom}
\end{align}

The influence of electron-phonon coupling on the phase transition, specifically on the term $\tau$, is of interest. Equation \eqref{condom} was rearranged to solve for $\tau$, yielding:

\begin{align}
\tau &\leq \sqrt{\frac{2}{N_\text e}\left (\frac{\hbar \omega}{\Delta}\right )^2 +\frac{1}{4}}+\frac{1}{2} & \tau &\geq \sqrt{\frac{2}{N_\text e}}\frac{\hbar \omega}{\Delta} \label{cond-kern}
\end{align}

Equation \eqref{cond-kern} shows that the presence of the nucleus reduces the minimum coupling strength required for the phase transition. In addition, it should be noted that there is an upper limit for the strength of electron-phonon coupling. If the coupling becomes too strong, the phase transition will not occur, and the nucleus will remain shifted. This situation, which is characterized by strong electron-phonon coupling, is referred to as a polaron. Moreover, the inverse Curie temperature can be calculated as
\begin{align}
\beta_{\text C}=\frac{2}{\Delta}\arctanh\left( \tau-2\left(\frac{\hbar \omega}{\Delta} \right)^2\frac{1}{N_\text e\tau} \right) \label{TcOM}
\end{align}

\begin{figure}
    \includegraphics[width=0.45\linewidth]{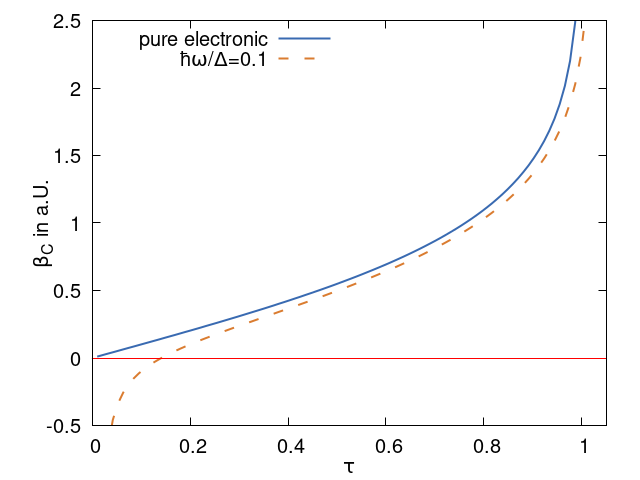}
    \includegraphics[width=0.45\linewidth]{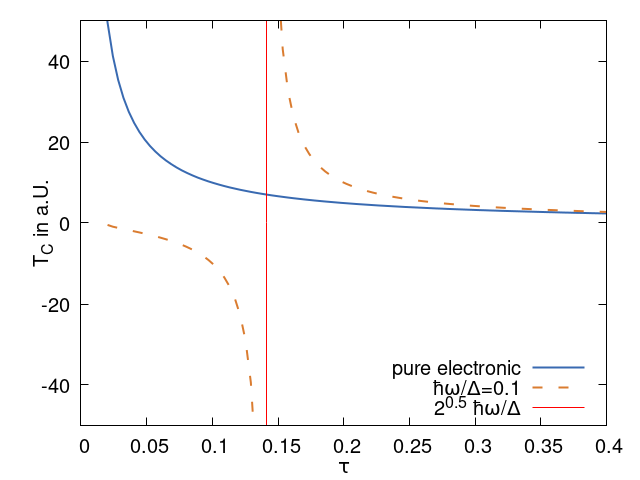}
    \caption{The impact of nucleus corrections on the Curie temperature is depicted in the provided plots. The plot on the left illustrates the dependence of the inverse Curie temperature on the parameter $\tau$, which is represented in arbitrary units. On the right, the corresponding Curie temperature values are displayed. The graphs showcase the influence of the corrections under different conditions of bandgap and TO modes, and are contrasted with the results obtained without considering nucleus corrections, as indicated in the plot. As $\tau$ decreases, the significance of the corrections becomes more pronounced.}\label{tempsh}
\end{figure}

When comparing Equation \eqref{TcOM} with the pure electronic case Equation \eqref{Tc}, it is observed that the nucleus correction term $2\left(\frac{\hbar \omega}{\Delta} \right)^2\frac{1}{N_\text e \tau}$ is present, leading to a minor correction. Consequently, the lower temperature phase is stable at higher temperatures than predicted in the pure electronic case. However, if the electron-phonon coupling becomes too strong, no phase transition occurs. This finding is demonstrated in \textbf{Figure \ref{tempsh}}
The nucleus correction has a significant impact on the material's behavior, as shown in \textbf{Figure \ref{tempsh}}. Unlike the pure electronic result, the nucleus correction enables phase transitions even for $\tau>1$. This suggests that the influence of the nucleus boosts the electron-phonon coupling. \\
To test the nucleus-correction, we use the example from KK\cite{vibronicth}. They look at the movement of the Ti in BaTiO$_3$. The Ti mass is 47.87 g/mol. The frequency is $\omega=0.8\cdot 10^{13}\text{s}^{-1}$, the electron-phonon coupling constant is $V=1.2$eV/Å, and the used band gap is $\Delta=5$eV. In their example, they set $N_\text{e}=N=1$. Using the properties, $\tau$ is calculated as 0.552 and the correction term for the phase transition temperature as $2\left(\frac{\hbar \omega}{\Delta} \right)^2\frac{1}{N_\text e\tau}=4\cdot 10^{-6}$, which yields in a minimal contribution.


\section{Conclusion} \label{Conclusion}
In this paper, we conducted a comprehensive review of the main findings presented by Kristoffel and Konsin, carefully validating and completing their outcomes. We recognized the need to provide a detailed explanation of every approximation made and meticulously developed the procedure, considering the scarcity of literature in this area.

An important aspect of our study was the incorporation of the second Bogoliubov inequality, which allowed us to introduce nucleus correction terms into the model. These corrections have the potential to significantly influence the behavior of phase-change materials and provide criteria for the formation of polarons. 

Overall, our work contributes to the advancement of the field by clarifying existing models, introducing new concepts, and outlining promising avenues for future research, similar to the work of Raffaello Bianco et al. \cite{PhysRevB.96.014111} and the currently work of Marios Zacharias et al. \cite{PhysRevB.108.035155}.


\medskip
\textbf{Acknowledgements} \par 
This work has been performed through funding, which we gratefully acknowledge,  by the EC FET-Open project PHEMTRONICS under the Horizon 2020 grant agreement No 899598. It is a pleasure to thank Dr. Helga Böhm for stimulating discussionss, and Dr. Christoph Cobet for providing insight into experimental techniques and critical manuscript reading.

\medskip
\section{Appendix: The Electronic System}\label{Appendix_Electronic}
\setcounter{equation}{0}
\renewcommand{\theequation}{A.\arabic{equation}}
The energy of the electronic system, denoted as $\bar{\varepsilon}(X)$, can be computed using Equation \eqref{H_el}. It is important to note that the Hamiltonian in Equation \eqref{Ham} describes a many-electron system in which each electron interacts with one phonon, and thus the electronic Hamiltonian in Equation \eqref{H_el} must be expressed using the hole tensor formalism in order to fully account for electron entanglement. The Hilbert space basis for the electronic component is the one-electron Hilbert space $H_{1}$, and utilizing these spaces allows us to express the electronic Hamiltonian in Equation \eqref{H_el} as 
\begin{align}
\Ham_{\text{el,BO}}=&\sum_{\sigma}\varepsilon_{\sigma}\left(\underbrace{\ak_{\sigma}\av_{\sigma}\otimes\mathbb{1}_{H_1}\otimes\mathbb{1}_{H_1}\ldots\otimes\mathbb{1}_{H_1}}_{N_{\rm e}}+\mathbb{1}_{H_1}\otimes\ak_{\sigma}\av_{\sigma}\otimes\mathbb{1}_{H_1}\ldots\otimes\mathbb{1}_{H_1}+\ldots + \mathbb{1}_{H_1}\otimes \ldots \otimes \ak_{\sigma}\av_{\sigma}\right)\notag \\ +&\underbrace{\mathbb{1}_{H_1}\otimes\mathbb{1}_{H_1}\otimes\mathbb{1}_{H_1}\ldots\otimes\mathbb{1}_{H_1}}_{N_{\rm e}}\frac{M\omega^2}{2}X^2 \notag \\ +& \sum_{\sigma,\sigma'}\frac{V_{\sigma\sigma'}}{\sqrt{N}}\left(\underbrace{\ak_{\sigma}\av_{\sigma'}\otimes\mathbb{1}_{H_1}\otimes\mathbb{1}_{H_1}\ldots\otimes\mathbb{1}_{H_1}}_{N_{\rm e}}+\mathbb{1}_{H_1}\otimes\ak_{\sigma}\av_{\sigma'}\otimes\mathbb{1}_{H_1}\ldots\otimes\mathbb{1}_{H_1}+\ldots + \mathbb{1}_{H_1}\otimes \ldots \otimes \ak_{\sigma}\av_{\sigma'}\right) X
 \label{Htens}
 \end{align}
First, let's examine the solutions of the non-interacting part of the Hamiltonian in Equation \eqref{Htens}, neglecting the third line. We are looking for wavefunctions that are tensor products of the one-electron eigenfunctions $\varphi_1$ and $\varphi_2$, denoted by $\varphi_{i_1}\otimes \varphi_{i_2}\otimes \ldots \otimes \varphi_{i_{N_{\rm e}}}$. The corresponding energy eigenvalues of the non-interacting part of the Hamiltonian are given by $N_1\varepsilon_1+N_2\varepsilon_2$, where $N_1$ and $N_2$ are the occupation numbers of the electron bands. Note that the energy depends on the occupation of the electron bands. The wavefunctions satisfying this equation are given by the following tensor product:
\begin{align}
\sum_{\sigma}\varepsilon_{\sigma}&\left(\ak_{\sigma}\av_{\sigma}\otimes\mathbb{1}_{H_1}\otimes\mathbb{1}_{H_1}\ldots\otimes\mathbb{1}_{H_1}+\ldots + \mathbb{1}_{H_1}\otimes \ldots \otimes \ak_{\sigma}\av_{\sigma}\right) \varphi_{i_1}\otimes \varphi_{i_2}\otimes \ldots \otimes \varphi_{i_{N_{\rm e}}}= \\ & \left(N_1\varepsilon_1+N_2\varepsilon_2\right) \varphi_{i_1}\otimes \varphi_{i_2}\otimes \ldots \otimes \varphi_{i_{N_{\rm e}}} 
\label{eq:wavefunction}
\end{align}
where there are $2^{N_{\rm e}}$ different tensor wavefunctions with the corresponding total energies. In this case, we have degenerate energies, and the number of degenerate states is given by $\binom{N_{\rm e}}{N_1}$, as shown in Equation \eqref{entar}
 \begin{equation}
    \#\left(N_1\varepsilon_1+\left(N_\text e-N_1\right)\varepsilon_2\right)=\binom{N_{\rm e}}{N_1} \label{entar}
    \end{equation} 
    As we have seen every electronic Hamiltonian commutes with the full Hamiltonian. Because of the form of the interaction the same argument holds for Hamiltonian \eqref{Htens}. This leads to a much simpler system of equations to solve
\begin{subequations} \label{com_rel}
    \begin{align}
         \Ham_{\text{el,BO}}\Phi_{n_1 \ldots n_{N_\text e}}(X)=&\left (\varepsilon'_{n_1}(X)+\ldots \varepsilon'_{n_{N_\text e}}(X)+\frac{M\omega^2X^2}{2}\right )\Phi_{n_1 \ldots n_{N_\text e}}(X)=\bar{\varepsilon}(X)\Phi_{n_1 \ldots n_{N_\text e}}(X) \label{total_E}  \\ &\left(\sum_{\sigma}\varepsilon_{\sigma}\ak_{\sigma}\av_{\sigma}+\sum_{\sigma,\sigma'}\frac{V_{\sigma\sigma'}}{\sqrt{N}}\ak_{\sigma}\av_{\sigma'}X\right )\Phi_{n}(X)=\varepsilon'_{n}(X)\Phi_{n}(X) \label{e_bar}
    \end{align}
\end{subequations}
Due to the commutator relations we know that 
\begin{equation}
    \Phi_{n_1 \ldots n_{N_\text e}}(X)=\Phi_{n_1}(X)\otimes\Phi_{n_2}(X)\otimes\ldots\Phi_{n_{N_e}}(X)
\end{equation}
 The desired energy in Equation \eqref{total_E} one obviously finds as
 \begin{align}
     \bar\varepsilon(X)=\varepsilon'_{n_1}(X)+\ldots \varepsilon'_{n_{N_\text e}}(X)+\frac{M\omega^2X^2}{2}
 \end{align}
 Solving Equation \eqref{e_bar} is also analytically possible. Using the ansatz 
 \begin{equation}
     \Phi_{n}(X)=c_{1}(X)\ \varphi_{1}+c_{2}(X)\ \varphi_{2}
 \end{equation}
 one find the system of equations
  \begin{align}
 \mqty( \varepsilon_1 & \frac{V}{\sqrt{N}}X \\ \frac{V}{\sqrt{N}}X &\varepsilon_2 )\mqty(c_{1}(X) \\  c_{2}(X))=\varepsilon'_{n}(X)\mqty(c_{1}(X) \\  c_{2}(X)) \label{1T}
  \end{align}
The solution of Equation \eqref{1T} yields the desired energies
 \begin{align}
  \varepsilon'_{n}(X)=\varepsilon'_{\mp}(X)=\frac{\varepsilon_1+\varepsilon_2\mp \sqrt{\Delta^2+4\frac{V^2}{N}X^2}}{2}\label{renormen}
\end{align}
which can be interpreted as a renormalization of the bands $\varepsilon$ as found by \cite{Kristoffel1973}. The assosicated states are calculated as
\begin{equation}
   \ket{\mp}(X)\equiv\Phi_{n}(X)=\mp\frac{1}{\sqrt{2}}\sqrt{1 \mp \frac{\Delta}{E_{\rm g}(X)}}\varphi_{1}+\frac{1}{\sqrt{2}}\sqrt{1 \pm \frac{\Delta}{E_{\rm g}(X)}}\varphi_{2}  \label{renormen1}
\end{equation}
Here we introduced the bandgap without and with electron-phonon interaction:
\begin{align}
    \Delta&=\varepsilon_2-\varepsilon_1 & E_{\text g}(X)&=\varepsilon'_{+}(X)-\varepsilon'_{-}(X)
\end{align}
The electronic energy is straightforwardly calculated as
\begin{align}
    \bar\varepsilon_{N_1}(X)=N_1\varepsilon'_{+}(X)+(N_\text e-N_1)\varepsilon'_{-}(X)+\frac{M\omega^2}{2}X^2 \label{renormen2}
\end{align}
with the quantum number $N_1$ which is again degenerated with  
\begin{align}
    \#N=\binom{N_\text e}{N_1} \label{entartung}
\end{align}

\medskip

%
\bibliographystyle{MSP}
\bibliography{Literatur}


\end{document}